\documentclass[twocolumn,prl,superscriptaddress,showpacs,aps,10pt]{revtex4-1}
\newcommand{\pagenumbaa}{1}
\usepackage{graphicx}
\usepackage{color}
\usepackage{float}

\usepackage{hyperref}

\usepackage{verbatim}		

\usepackage{units}
\begin{document}


\title{Edge State Wave Functions from Momentum-Conserving Tunneling Spectroscopy}


\author{T.~Patlatiuk}\altaffiliation{These authors contributed equally to this work}
\affiliation{Departement Physik, University of Basel, Klingelbergstrasse 82, CH-4056 Basel, Switzerland}

\author{C.~P.~Scheller}\altaffiliation{These authors contributed equally to this work}
\affiliation{Departement Physik, University of Basel, Klingelbergstrasse 82, CH-4056 Basel, Switzerland}

\author{D.~Hill}
\affiliation{Department of Physics and Astronomy, University of California, Los Angeles, California 90095, USA}

\author{Y.~Tserkovnyak}
\affiliation{Department of Physics and Astronomy, University of California, Los Angeles, California 90095, USA}

\author{J.~C.~Egues}
\affiliation{Instituto de F\'isica de S\~ao Carlos, Universidade de S\~ao Paulo, 13560-970 S\~ao Carlos, S\~ao Paulo, Brazil}

\author{G.~Barak}
\affiliation{Department of Physics, Harvard University, Cambridge, Massachusetts 02138, USA}

\author{A.~Yacoby}
\affiliation{Department of Physics, Harvard University, Cambridge, Massachusetts 02138, USA}

\author{L.~N.~Pfeiffer}
\affiliation{Department of Electrical Engineering, Princeton University, Princeton, New Jersey 08544, USA}

\author{K.~W.~West}
\affiliation{Department of Electrical Engineering, Princeton University, Princeton, New Jersey 08544, USA}

\author{D.~M.~Zumb\"uhl}
\email{dominik.zumbuhl@unibas.ch}\affiliation{Departement Physik, University of Basel, Klingelbergstrasse 82, CH-4056 Basel, Switzerland}

\begin{abstract}
We perform momentum-conserving tunneling spectroscopy using a GaAs cleaved-edge overgrowth quantum wire to investigate adjacent quantum Hall edge states. We use the lowest five wire modes with their distinct wave functions to probe each edge state and apply magnetic fields to modify the wave functions and their overlap. This reveals an intricate and rich tunneling conductance fan structure which is succinctly different for each of the wire modes. We self-consistently solve the Poisson-Schr{\"o}dinger equations to simulate the spectroscopy, reproducing the striking fans in great detail, thus confirming the calculations. Further, the model predicts hybridization between wire states and Landau levels, which is also confirmed experimentally. This establishes momentum-conserving tunneling spectroscopy as a powerful technique to probe edge state wave functions.


\end{abstract}
\maketitle

\setcounter{page}{\pagenumbaa}
\thispagestyle{plain}

Edge states play a key role in materials with a non-trivial topology establishing a conducting boundary around a (nominally) insulating bulk in novel topological insulators as well as quantum (spin) Hall materials \cite{Tsui1982,Stormer1983,Mendez1984,Novoselov2005,Zhang2005,Konig2007a,Tsukazaki2007,Tsukazaki2010}.
Despite clear theoretical understanding, only few experiments could probe and resolve edge states in systems with steep, hard wall-like confinement potentials~\cite{Pfeiffer1990,Yacoby1996,Yacoby1997,Auslaender2002, Kang2000,Huber2005a,Grayson2007,Steinke2013,PatlatiukScheller2018}. In semiconductor heterostructures with typical gate defined or etched structures, strong confinement is difficult to produce, opening the door for Coulomb interactions to dominate and leading e.g. to issues like edge state reconstruction \cite{Chklovskii1992, Chamon1994a}. On the other hand, steep confinement can arise naturally by virtue of the sample fabrication, for example in van der Waals heterostructures \cite{Li2013} or in GaAs cleaved edge overgrowth heterostructures \cite{Pfeiffer1990,Yacoby1996,Yacoby1997,Auslaender2002,Kang2000,
Huber2005a,Grayson2007,Steinke2013,PatlatiukScheller2018}.

\begin{figure}[t]
\centering
\includegraphics[width=1\columnwidth]{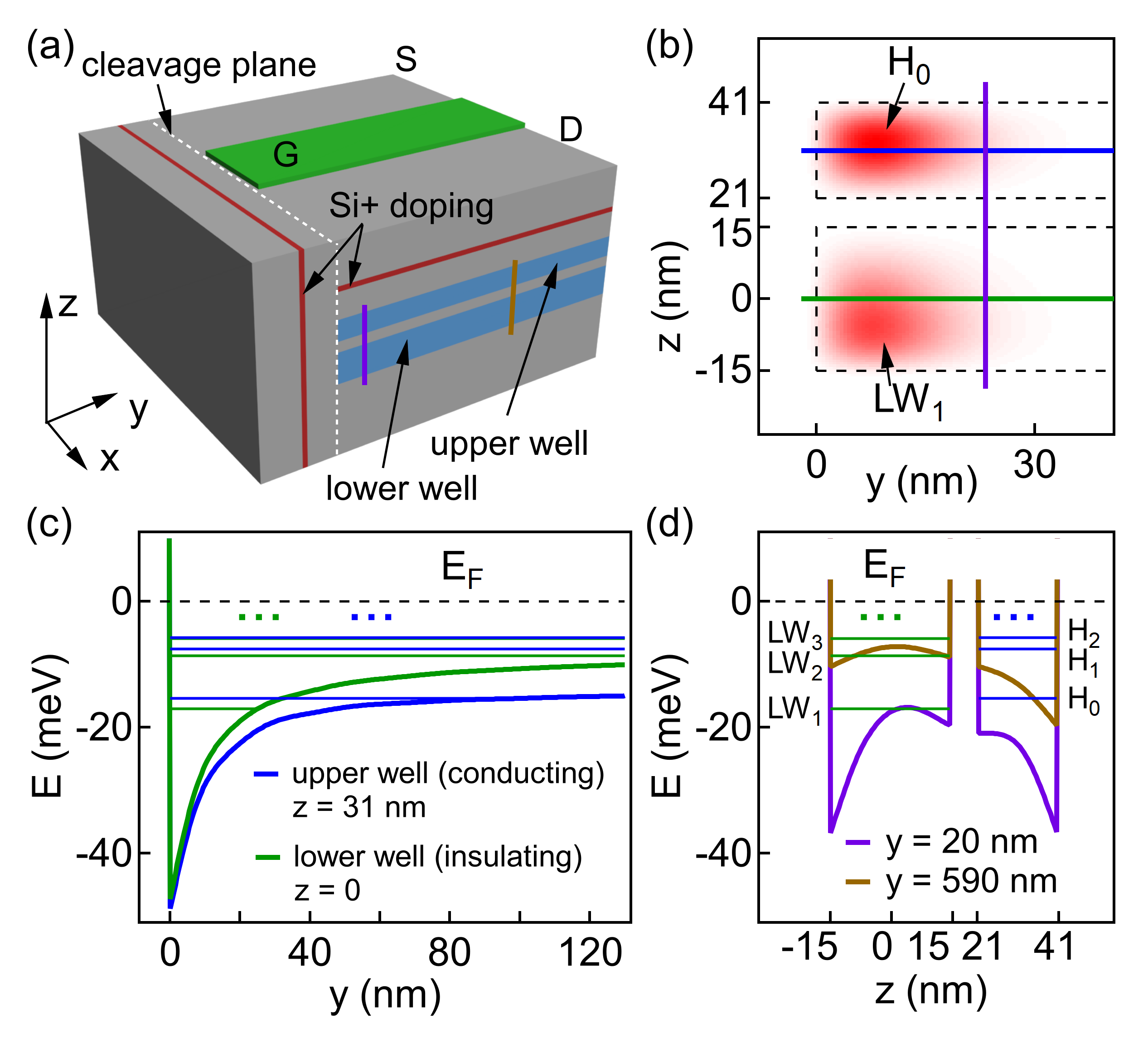}\vspace{-2.5mm}
\caption{(a)~Sample schematic showing the upper and lower quantum well (light blue) together with top and side Si dopants (red). A surface gate G (green) controls the local electron density. Source and drain ohmic contacts are labeled S and D, respectively. (b)~Wave function $\psi (y,z)$ for the ground state $H_0$ ($LW_1$) in the upper (lower) quantum well. (c)~Conduction band profiles along the y-direction in the upper (blue) and lower (green) quantum wells. Horizontal lines represent the lowest three eigenenergies. (d)~Conduction band profile along the z-direction 20 nm (purple) and 590 nm (brown) away from the cleavage plane (dashed white in (a)).}\vspace{-6mm}
\label{fig:1}
\end{figure}

The wave function profiles of edge states are very difficult to access in experiments and their properties are often inferred from standard transport measurements \cite{Alphenaar1990a,Roth2009,Brune2010}, which may also suffer from remnant bulk conductivity \cite{Konig2013}. Scanning probe techniques can offer valuable additional insight into localized states \cite{Tessmer1998a,Steele2005} as well as edge states both in quantum Hall (QH) \cite{Yacoby2003,Suddards2012b,Lai2011a,Pascher2014} and in quantum spin Hall \cite{Nowack2013,Konig2013,Ma2015,Shi2019} regimes. However, scanning probe methods also have a number of limitations, including poor resolution and invasive probes. Thus, establishing methods to directly access the edge state wave function is a great challenge.

\begin{figure}
\centering
\includegraphics[width=1\columnwidth]{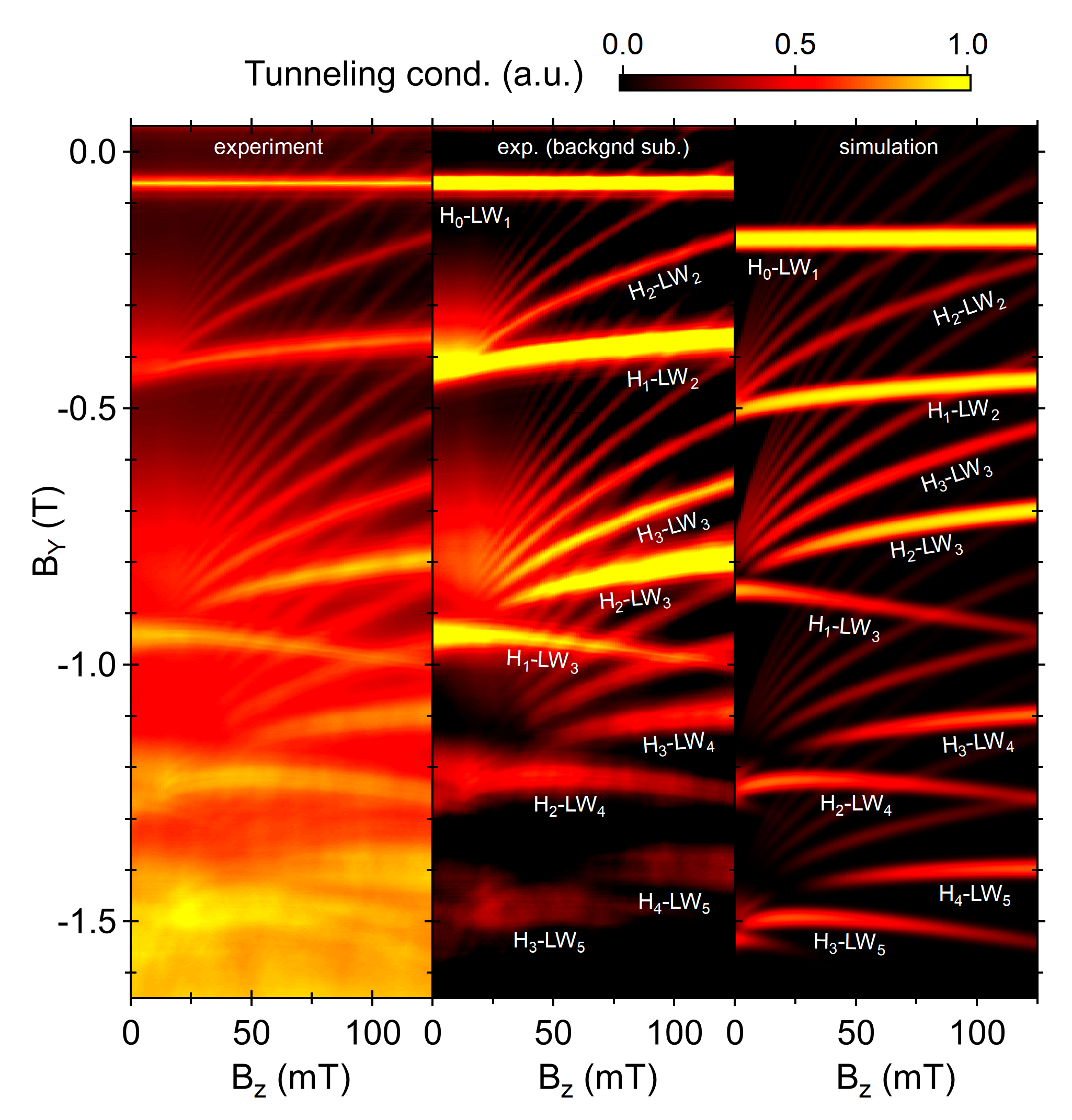}\vspace{-2.5mm}
\caption{Experimental (left and middle) and theoretical (right) tunneling conductance between various hybrid states and lower wire modes as a function of out-of-plane ($B_Z$) and in-plane ($B_Y$) magnetic fields. For the middle panel, a smooth background is subtracted from the experimental data and an oversaturated color scale is used to emphasize weak features. The tunneling conductance to the lower wire mode $LW_2$ is strongest for the hybrid state $H_1$ and depends weakly on the magnetic field $B_Z$. In contrast, the tunneling conductance to the wire modes $LW_{3,4,5}$ show significant strength variations as magnetic field $B_Z$ is increased, see main text for details.
}\vspace{-6mm}
\label{fig:2}
\end{figure}

\begin{figure}[t]
\includegraphics[width=1\columnwidth]{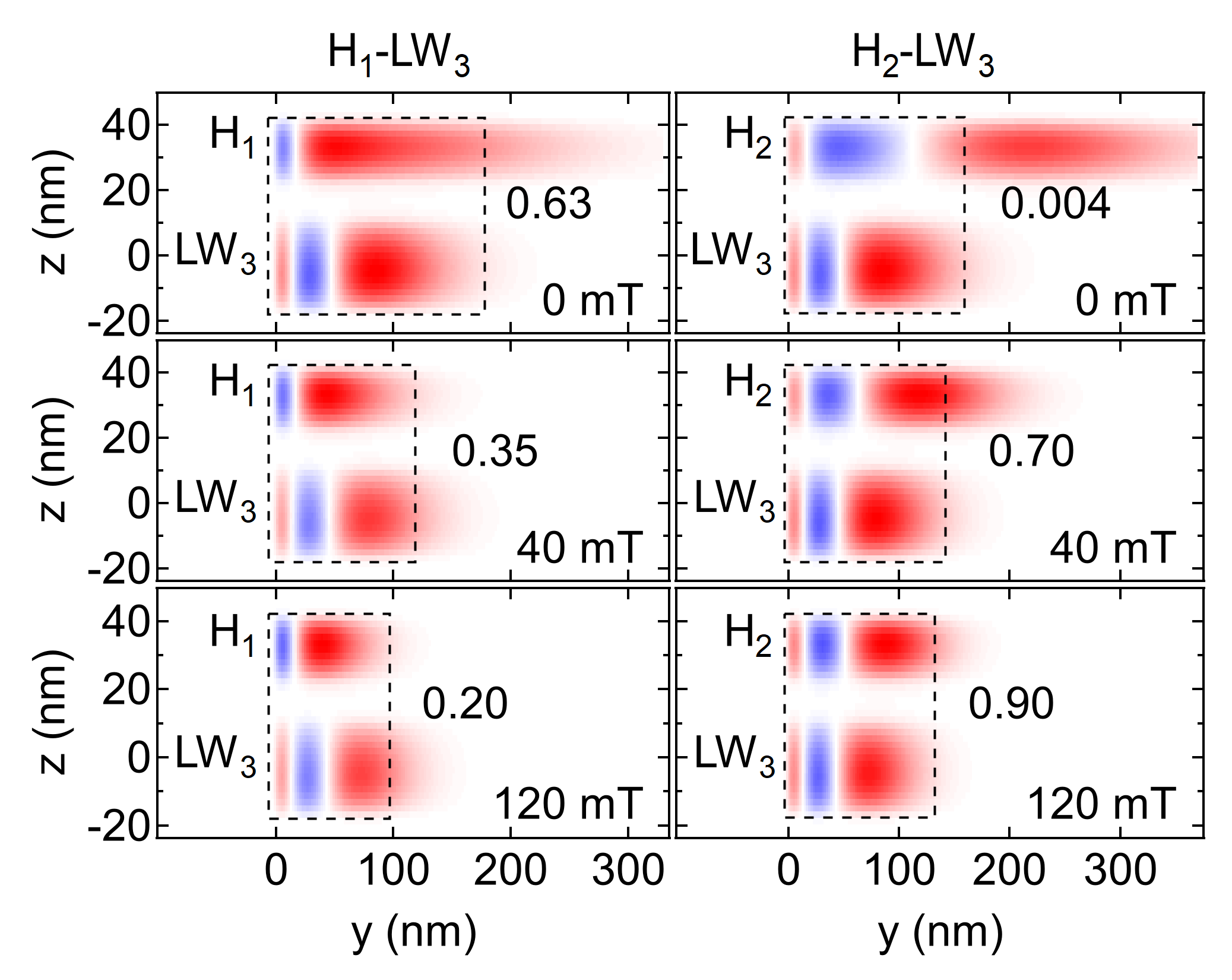}\vspace{-2.5mm}
\caption{Evolution of the wave functions of the hybrid state $H_1$ (left column), $H_2$ (right column), and lower wire mode ($LW_3$) as a function of magnetic field $B_Z$. The cleaved edge is at $y=0$. The wave function overlap between $LW_3$ and corresponding hybrid state, normalized to its maximal value in Fig.~\ref{fig:2}, is indicated as a numeric value in each panel. Red and blue colors represent the signs of the probability amplitude.
}\vspace{-4mm}
\label{fig:3}
\end{figure}

In this work, we use tunneling into modes of a cleaved-edge overgrowth quantum wire as an energy and momentum selective spectrometer to investigate the wave functions of QH edge states in an adjacent quantum well, see sample schematic in Fig.\,\ref{fig:1}(a). We measure the resonant tunneling conductance, related to the wave function overlap, as a function of in-plane magnetic field $B_Y$, which controls the relative momentum, and perpendicular magnetic field $B_Z$, which predominantly modifies the QH wave functions. This produces a set of fans of intricately growing and fading curves, see Fig.\,\ref{fig:2}. We self-consistently simulate the edge state and quantum wire wave functions, see Fig.\,\ref{fig:1}(b) and Fig.\,\ref{fig:3}, which allows us to calculate the measured tunneling conductance. We find very good agreement with experiment, thus confirming the validity of the calculated wave functions. Finally, as the QH edge states and wire states in the upper system are overlapping close to the edge, these two types of states are also hybridizing, leading to the formation of avoided crossings. These are also experimentally observed when brought into resonance with the lower wire, further confirming the simulations.

The schematic cross-section of the double quantum well sample used in this study is shown in Fig.\,\ref{fig:1}a. It contains upper and lower GaAs quantum wells (blue) separated by a thin AlGaAs tunnel barrier (gray). A doping layer (red) above the upper well provides charges for a high-mobility upper 2DEG, while electron density in the lower well is below the conduction threshold, making it insulating, as if there was no lower 2DEG. Dopants deposited above an in-situ cleaved surface \cite{Yacoby1996} (vertical red) create a quasi-triangular confinement potential forming extended 1D wire modes in both upper and lower wells. We name these modes $LW_1, LW_2, ...$ in the lower system. In the upper system, the 1D wire modes hybridize with the Landau levels present at finite $B_Z$ and form what we refer to as hybrid states $H_0, H_1, ...$. In this double quantum well sample, momentum selective spectroscopy allows us to use the lower wire modes $LW_{1,2,...}$ to probe the upper system hybrid states $H_{0,1,...}$. Source and drain ohmic contacts are attached far away from the cleavage plane, see Ref.~\onlinecite{PatlatiukScheller2018} for further sample details.

Electrical conductance between the source (S) and drain (D) was measured as a function of magnetic field $\textbf{B}=(B_Y,B_Z)$, orientations are shown in Fig.\,\ref{fig:1}a. To tune the double well device into the tunneling regime, a negative voltage was applied to the surface gate G (green) locally depleting the 2DEG and the upper wire modes. In this configuration, electrons from the source ohmic contact propagate through the 2DEG region, tunnel to the lower wire, propagate along the lower wire under the gate and tunnel to the 2DEG connected to the drain on the other side of the gate. All the measurements presented in this paper were done with a small AC excitation (lock-in technique) at zero DC bias, ensuring that tunneling takes place only between states close to the Fermi level. Microwave filters and thermalizers \cite{Scheller2014} on each lead provide an electron temperature around 10\,mK, see \cite{PatlatiukScheller2018} for details.

Due to translational invariance in the tunneling region away from the gate, electron momentum $k_x$ is conserved during the tunneling between the upper and lower systems, giving resonant tunneling when states in the upper and lower system have matching momenta. The in-plane magnetic field $B_Y$ changes the electron momenta by $\Delta k_x=e d_Z B_Y /\hbar$, where $d_Z$ is the z displacement between the center of mass of initial and final states. Similarly, the perpendicular magnetic field $B_Z$ provides another contribution to the momentum shift $\Delta k_x=e d_Y B_Z /\hbar$ for states displaced by $d_Y$ in the $y$ direction and also sets the Landau level energies.

The resulting spectroscopy using both magnetic fields $B_Y$ and $B_Z$ displays several intricate fan structures of rising and decreasing curves, as shown in Fig.~\ref{fig:2}. Each fan arises from tunneling to one of the lower wire modes $LW_2,LW_3,...$. Each curve within a fan corresponds to a hybrid state (upper system). The fans are shifted in $B_Y$ to compensate the mismatch in Fermi momentum between the 2D gas and the different lower wire modes \cite{PatlatiukScheller2018}. The left panel of Fig.~\ref{fig:2} displays the measured differential conductance with magnetic fields rotated into the sample coordinate system (see Fig.~\ref{fig:1}a)). On the middle panel, a smooth background was subtracted to emphasize the peak structure. The right panel shows the simulated results.

The horizontal line close to $B_Y = 0$ corresponds to tunneling between the lowest modes in the upper ($H_0$) and lower ($LW_1$) systems. The top most fan originates from tunneling into $LW_2$, shows the strongest signal for tunneling from the state $H_1$ and weaker tunneling for the states with higher orbital indices ($H_2$, $H_3$\,...). The second fan results from the tunneling into $LW_3$, shows substantial decrease of the tunneling signal for $H_1$ as the magnetic field $B_Z$ increases and opposite behavior for $H_2$. The other curves of this fan have a maximum at intermediate values of $B_Z$ accompanied by small signal at low and high fields. The fan originating from tunneling into $LW_4$ is very similar to the fan $LW_3$ with the difference that the indices of all the curves of this fan are shifted up by one and the first curve disappears completely already at small fields. Similar behavior is observed for the fan with tunneling into $LW_5$, showing decreasing signal for $H_3$ and increasing signal $H_4$, and again, very strong suppression of the lower states, here $H_0$, $H_1$, and $H_2$.

To explain this rich and striking pattern of the tunneling conductance, we have numerically calculated the wave functions for the states in the upper and lower systems using a 2D self-consistent Schr\"odinger-Poisson solver \cite{RotherM}. The wave function of an electron is written as a product of a plane wave with momenta $k_x$ and the self-consistent solutions in y-z plane:
\begin{equation}
\Psi_n^j (x,y,z,k_x) = e^{i k_x x} \psi_n^j (y,z).
\label{equ:wave_fucntions}
\end{equation}
The index $n$ enumerates different orbital states, $j=u$ denotes the upper system, and $j=l$ the lower system. Figure \ref{fig:1}b shows the $\psi_0^u (y,z)$ and $\psi_1^l (y,z)$ components of the wave functions $H_0$ and $LW_1$. The conduction band along the y- and z-directions are shown in Fig.\,\ref{fig:1}c and d, respectively. A decrease of the conduction band energy close to the cleavage plane at $y=0$ is caused by the electric field of the overgrown ionized dopants nearby at $y<0$. The resulting triangular confinement potential in Fig.\,\ref{fig:1}c is stronger for the lower well leading to the presence of several lower wire modes. The energies of the lowest three states are depicted as blue and green horizontal lines for upper and lower system, respectively.

The Landau gauge $\textbf{A}=(zB_Y-yB_Z,0,0)$ was used to describe the system at finite magnetic field $\textbf{B}$, as it captures the translational invariance along the $x$ direction. In this gauge, momentum $k_x$ is a good quantum number, so the total wave function $\Psi_n^j (x,y,z,k_x,\textbf{B})$ can be written analogously to Eq.\,\ref{equ:wave_fucntions}.
Here, $\psi_n^j$ was calculated under the assumption that at a finite magnetic field, the Fermi level and electrostatic confinement potential are the same as at $\textbf{B}=0$. This assumption is justified as small magnetic fields do not modify the electron density much.

Using Fermi's golden rule, the tunneling conductance between state $H_n$ in the upper and state $LW_m$ in the lower systems can be written as:
\begin{equation}
G_{n \to m} = \frac{2\pi e}{\hbar}t^2 \rho_u\rho_l \lambda_{n \to m}^2
\label{equ:tun_current}
\end{equation}
where $e$ is electron charge, $\hbar$ is the Plank constant, $t$ is the tunneling coupling strength, $\rho_j$ is the density of states, and $\lambda_{n \to m}$ is the wave function overlap between the $\Psi_n^u$ and $\Psi_m^l$ states. An additional series resistance due to the lower wire under the gate was included. The calculated tunneling conductance into lower wire modes $LW_1 - LW_5$ is displayed in the right panel of Fig.\,\ref{fig:2}, giving striking agreement to the experiment. The observed conductance evolution can be explained by a simple selection rule. If the number of wave-function lobes in the tunneling region is the same for the initial and final states, then the conductance is strong. Otherwise, it is weak due to near orthogonality. The tunneling region is defined as the region in space where the wave functions in the upper and lower systems have a significant overlap (see Supplemental Material \cite{SOM}).

As an example, we will focus on the $H_2-LW_3$ resonance. The simulated wave functions of states $H_2$ and $LW_3$ at $B_Z = 0$ are shown in the top right panel of Fig.\,\ref{fig:3}. The tunneling region, shown as a dashed box, is limited by the smaller wave function, in this case $LW_3$. For these two states, the number of lobes in the tunneling box at zero field differs by one, resulting in small overlap and correspondingly low tunneling conductance. This is consistent with the measured small conductance of the $H_2-LW_3$ resonance at small $B_Z$ field in Fig.\,\ref{fig:2}. As the magnetic field is increased (Fig.\,\ref{fig:3} lower right panels) the already electrostatically more confined wire state $LW_3$ is hardly affected while the more spread out hybrid state $H_2$ is compressed and thus the third lobe starts to enter the tunneling region. The overlap between the wave functions thus grows with field, consistent with the observed increase in tunneling conductance for this resonance in both simulation and experiment in Fig.\,\ref{fig:2}.

There are some exceptions from this rule, which can be explained by a dominant contribution to the overlap from the outermost lobes, those most removed from the hard wall of the cleaved edge. Due to the quasi-triangular shape of the confinement potential, the outermost lobes of the wave function have by far the most weight. This is the case for the tunneling resonance $H_1-LW_3$, shown in the left column of Fig.~\ref{fig:3}. The selection rule predicts a weak tunneling conductance at zero field, as the number of lobes in the tunneling region (dashed box) is different for upper and lower wells. However, the actual overlap between these two wave functions is strong. This is due to the large overlap between the two last lobes, which overpowers the negative contributions from the other lobes. As the magnetic field $B_Z$ increases, the magnetic compression of the edge states reduces the width of the last lobe and the selection rule holds again.

The amount of top and side dopants in the structure strongly affects the $B_Y$ position of the resonances and here, we adjusted both doping levels to match the positions of the simulated $H_n-LW_2$ and $H_n-LW_3$ resonances with experiment. Only the resonances $H_n-LW_4$ and $H_n-LW_5$ were slightly shifted down by hand in Fig.~\ref{fig:2} to agree with experiment. We note that the doping levels do not qualitatively affect the tunneling strength and the $B_Z$ dependence. The excellent agreement between the single particle theory and experiment covering numerous striking and complicated features, see Fig.~\ref{fig:2}, is a very strong indication of the validity of the calculated wave functions.

\begin{figure}[t]
\includegraphics[width=1\columnwidth]{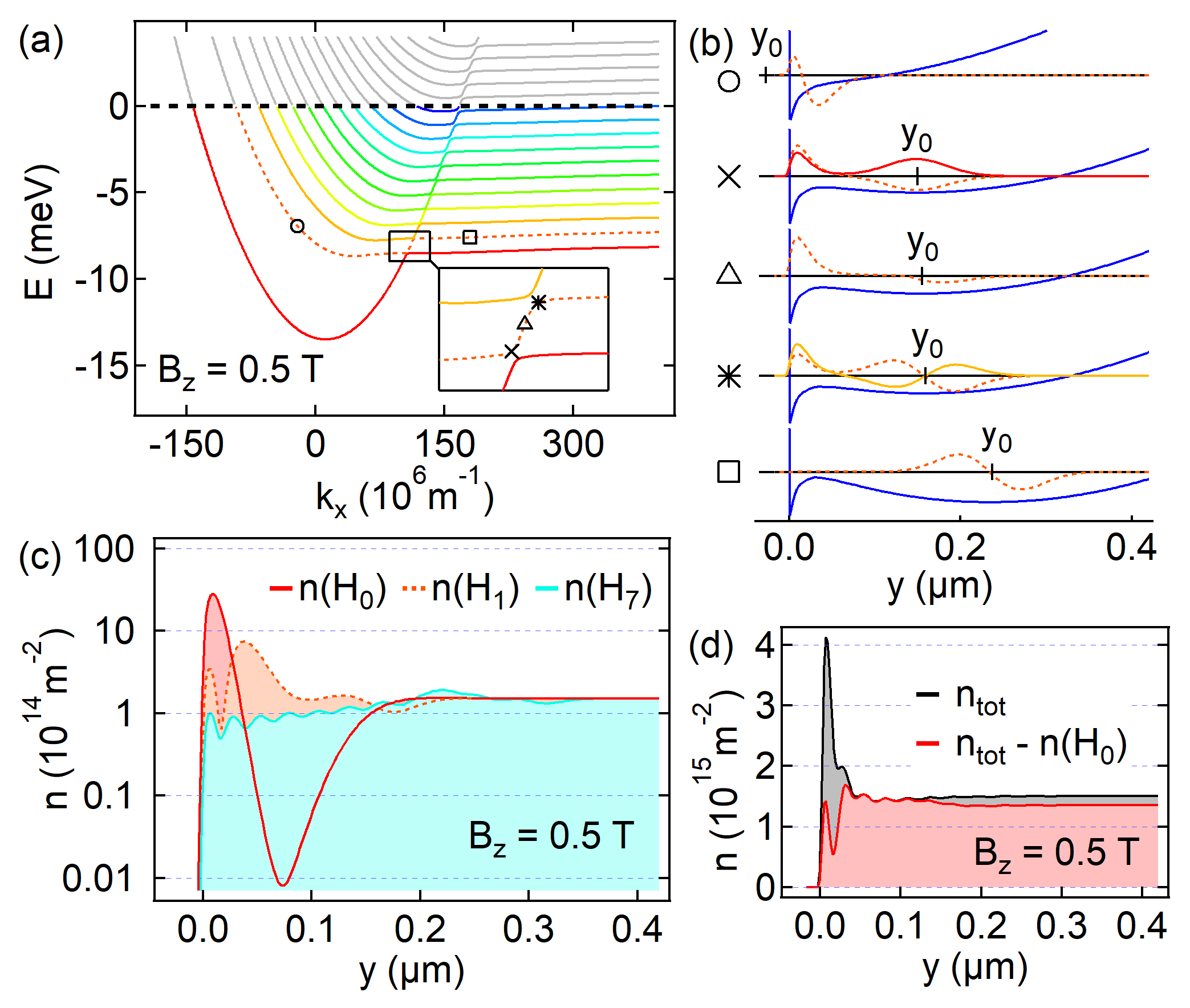}\vspace{-2.5mm}
\caption{(a) Dispersion of hybrid states ($H_0, H_1, ...$, colors)  in the upper system. Empty states above the Fermi level are shown in gray. (b) The wave function of the hybrid state $H_0$ (red), $H_1$ (dashed orange), $H_2$ (solid orange), and the total electrostatic confinement potential (blue) at $z=31$\,nm in the upper system for a range of values of the guiding center position $y_0$. The corresponding momenta $k_x$ are indicated with markers in (a). (c) Electron density of several hybrid states as well as (d) the total electron density with (black) and without (red) the contribution from $H_0$, as a function of distance $y$ from the cleavage plane, on the same horizontal scale as (b).
}\vspace{-6mm}
\label{fig:4}
\end{figure}

The triangular potential from the ionized overgrowth donors combined with the parabolic magnetic field confinement leads to a $k_x$-momentum dependent hybridization of the wire and quantum Hall states in the upper systems, shown in Fig.\,\ref{fig:4}a) and b). We note that $k_x$ is linked to the guiding center position $y_0$ of the parabolic magnetic confinement via $y_0 = k_x l_B^2$, with magnetic length $l_B=\sqrt{\hbar/eB_Z}$. Going through the $k_x$-axis, three distinct regimes separated by two anticrossings can be distinguished for each hybrid state (except for $H_0$, which undergoes only one anticrossing). Here, we pick the hybrid state $H_1$ as an example, with dispersion shown in Fig.~\ref{fig:4}a) and wave functions in b), both in dashed orange.

First, far away from the edge, corresponding to large $k_x$-momentum (square markers), this state has essentially the wave function of the second Landau level in the bulk. As $k_x$ is lowered, the state anticrosses with the hybrid state $H_2$ (stars), where the wave function hybridizes and delocalizes over the two minima in the potential (blue). The state $H_2$ in this anticrossing is the antibonding state with antisymmetric combination, shown in solid orange, featuring one additional node.

Second, for a small range of momenta $k_x$, the wave function of the state $H_0$ resembles that of the lowest wire mode (triangles), though still with one node and a very small negative amplitude in the parabolic minimum (shown amplified here for clarity). For even smaller momenta, the state undergoes a second hybridization (cross), now with hybrid state $H_0$, red. In a similar way, the upper wire mode hybridizes with all Landau levels, weaving through numerous anticrossings, which together are forming the parabolic wire dispersion at positive momenta, as seen in Fig.~\ref{fig:4}a. At each anticrossing, the wave function acquires one additional node when moving up one state in energy.

Third and finally, for negative $k_x$ (circle), the wave function is pushed more against the wall and more strongly confined by the magnetic field, thus increasing the energies of the states (Fig.~\ref{fig:4}a), still exhibiting one node, now corresponding to the second wire mode in the upper system.

The density profile of the hybrid states, shown in Fig.~\ref{fig:4}c), as well as the total electron density, shown in Fig.~\ref{fig:4}d), can be constructed from the wave functions by integrating over the momenta and $z$. Sufficiently far away from the edge, the hybrid states reach the bulk Landau level density, as expected. Close to the edge, a strong overshoot in density appears, due to the triangular wire confinement potential giving extra charges predominantly in $H_0$. The subsequent oscillations of density, most pronounced for $H_0$, are caused by the potentials forming an effective barrier between two minima.

\begin{figure}[t]
\includegraphics[width=1\columnwidth]{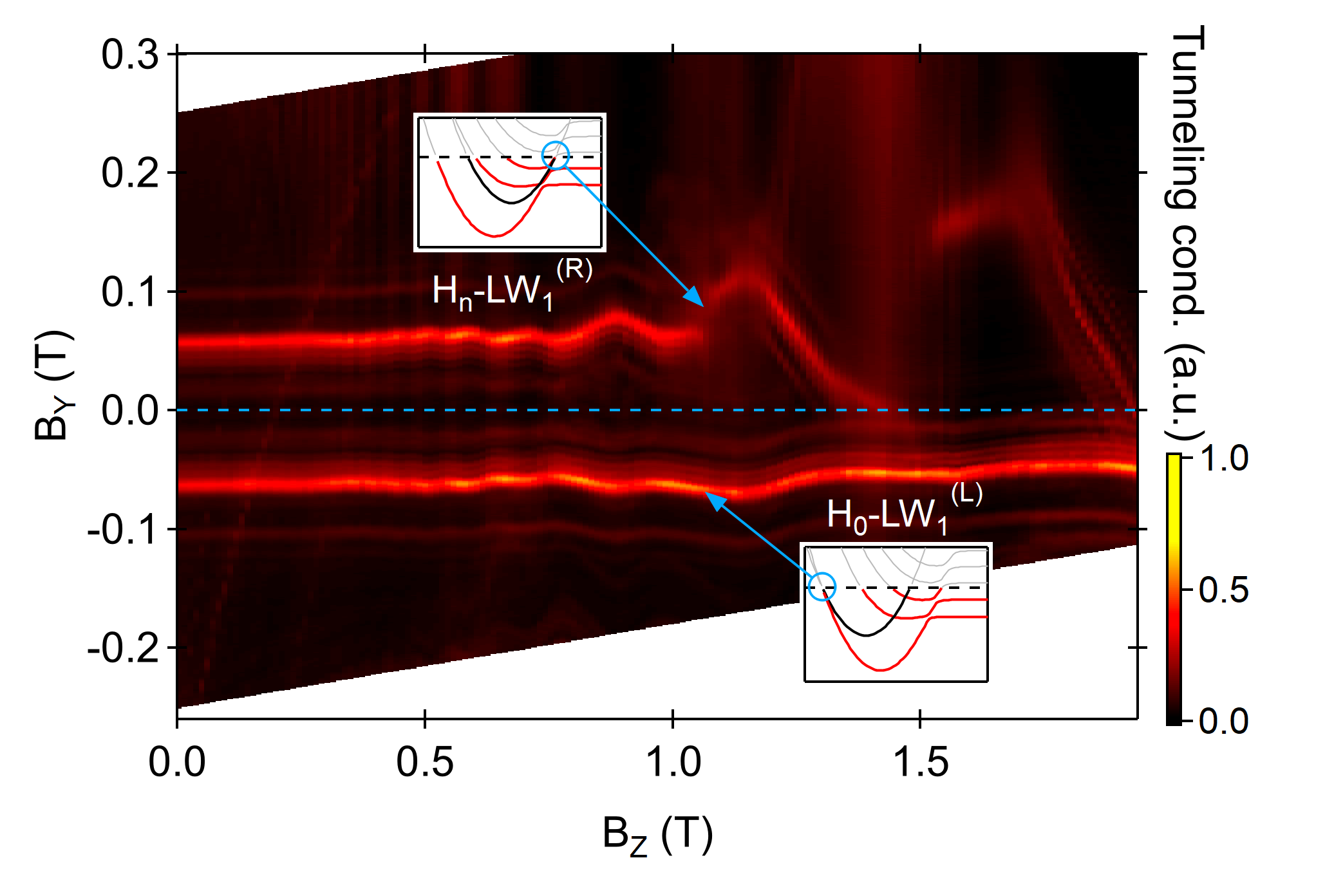}\vspace{-6mm}
\caption{Tunneling conductance between hybrid states in the upper system and the lower wire mode $LW_1$. Insets qualitatively depict the dispersion relation of the hybrid states (red) and $LW_1$ (black) for $B_Y$ and $B_Z$ which satisfy the resonant tunneling condition. Measurements extended for $B_Z<0$ are shown in Fig.\,S10 in the Supplemental Material \cite{SOM}.
}\vspace{-4mm}
\label{fig:5}
\end{figure}

Signatures of the hybridization as discussed above are present in the tunneling conductance, see Fig.~\ref{fig:5}, shown over a wider range of magnetic field $B_Z$ compared to Fig.~\ref{fig:2}. Insets qualitatively depict the dispersion relations of the hybrid states (red) and $LW_1$ mode (black) for the resonances indicated by the arrows. The resonance at negative $B_Y$ originates from the tunneling into the states close to the left Fermi point of the lower wire $LW_1^{(L)}$ (blue circle, lower inset). This resonance remains mostly unchanged in $B_Y$ as the magnetic field $B_Z$ is increased, giving a nearly horizontal feature as long as the wire confinement is dominant (for the full $B_Z$ range of Fig.~\ref{fig:5}). 
In contrast, the resonance originating form the right Fermi point $LW_1^{(R)}$, tracks the anticrossings of the hybrid states (blue circle, upper inset), leading to strong motion in $B_Y$. As the magnetic field $B_Z$ is increased, the topmost hybrid state is moving through the Fermi level, going through the regimes described in Fig.\,\ref{fig:4}b). This results even in some discontinuities when an avoided crossing is passing the Fermi level and the momentum is jumping to large $k_x$ to the next lower hybrid, as seen in Fig.~\ref{fig:5}.

In summary, we measured the tunneling conductance between the QH edge states and the lower wire mode as a function of both in-plane and out-of-plane magnetic fields. We show that the tunneling conductance calculated using numerically simulated wave functions reproduces the intricate fan structures with numerous striking features observed in experiment, giving a strong indication of the validity of the simulated wave functions. We formulate a simple selection rule to estimate the resonance strength based on wave function overlap or orthogonality in the tunneling region. The density at the cleaved edge is about a factor of three higher than in bulk due to the presence of the wire mode in the upper system. The simulation also provides the dispersions exhibiting hybridization between the wire mode and the Landau levels with numerous anticrossings. Upon inspection of the corresponding magnetic field range we indeed observe the predicted jumps of the resonances due to the anticrossings, thus providing further confirmation of the simulation.

The method used in this study could be also applied to probe the wave functions of spin split edge states or of fractional QH edges present at high magnetic fields. Additionally, samples with a side gate could be used to directly study influence of the confinement potential on the wave function of the QH edge and wire states. Finally, similar tunneling spectroscopy from a 1D conductor such as a nanowire or nanotube could also be used to probe edge states of topological insulators or other edge state materials of interest. More work would be required to understand the comparatively large broadening of the $LW_{4,5}$ resonances, or to capture better the magnetic field $B_Z$ where orthogonality to $LW_{3,4,5}$ is observed.

\begin{acknowledgments}
This work was supported by the Swiss Nanoscience Institute (SNI), NCCR QSIT, Swiss NSF No. 179024, ERC starting grant (DMZ), the EU H2020 European Microkelvin Platform EMP, grant No. 824109, Brazilian Grants No. 2016/08468-0 and No. 2016/50200-4 (SPRINT program), S\~ao Paulo Research Foundation (FAPESP). \\


\end{acknowledgments}


%

\end{document}